%% file: main.tex
\documentclass[prx,reprint,nofootinbib]{revtex4-2}
\usepackage{amsmath,amssymb,amsthm,braket,booktabs,graphicx}
\usepackage[colorlinks,citecolor=blue]{hyperref}
\usepackage{orcidlink}

\makeatletter
\newcommand\runinhead[1]{#1---}
\renewcommand\section{\@startsection{section}{1}{\parindent}{1ex}{0pt}{\itshape\runinhead}}
\renewcommand\@seccntformat[1]{\csname the#1\endcsname.\hspace{.3em}}
\makeatother

\def \be#1\ee {\begin{equation}#1\end{equation}}
\def \ba#1\ea {\begin{align}#1\end{align}}
\newcommand{\nn}{\nonumber}
\newtheorem{theorem}{Theorem}

\newtheorem{corollary}{Corollary}

\NewDocumentCommand{\fig}{m o}{\hyperref[fig:#1]{Figure~\ref*{fig:#1}\IfValueT{#2}{(#2)}}}
\newcommand{\tab}[1]{\hyperref[tab:#1]{Table~\ref*{tab:#1}}}
\newcommand{\thm}[1]{\hyperref[thm:#1]{Theorem~\ref*{thm:#1}}}
\newcommand{\lem}[1]{\hyperref[lem:#1]{Lemma~\ref*{lem:#1}}}
\newcommand{\cor}[1]{\hyperref[cor:#1]{Corollary~\ref*{cor:#1}}}
\renewcommand{\sec}[1]{\hyperref[sec:#1]{Section~\ref*{sec:#1}}}
\newcommand{\app}[1]{\hyperref[app:#1]{Appendix~\ref*{app:#1}}}

\begin{document}

\title{Certification failure in variational-principle observables}

\author{Ganga Singh Manchanda\,\orcidlink{0009-0009-8891-4766}\,}
\email{ganga.manchanda@chem.ox.ac.uk}
\affiliation{Physical and Theoretical Chemistry Laboratory, University of Oxford, United Kingdom}
\date{\today}


\begin{abstract}
Algorithms which prepare states via the variational principle assume that closeness in energy implies closeness in state; we formalise how this assumption can fail by showing that a ground-state energy estimate converged to within $\epsilon$ of the true value is compatible with observable expectation values in error by up to $\sqrt{2\epsilon\chi_O}$ at leading order, where $\chi_O$ is the static susceptibility of the observable. As a result, the observable error can be orders of magnitude larger than the energy error, a mechanism we call non-linear error amplification, and because a variational algorithm cannot guarantee the correctness of these properties from energy alone, energy does not necessarily certify an observable. In light of our findings, we examine three algorithms: the variational quantum eigensolver on the transverse-field Ising chain, the density-matrix renormalisation group on a pair of weakly coupled hydrogen chains, and Krylov quantum diagonalisation on lithium fluoride. In each case we demonstrate that a well-converged energy can leave observables uncertified.
\end{abstract}

\maketitle

\section{Introduction}
\label{sec:intro}

Understanding the behaviour of molecules, materials, and many-body systems is a central aim of the physical sciences. As a result, various methods, both classical and quantum, exist to compute the ground-state properties of a system. A large number of these approaches rely on the variational principle, which takes the energy of the state as a cost function to minimise in order to reach an approximation of the ground-state. An algorithm which achieves an energy within some small resolution $\epsilon$ is assumed to have obtained a well-approximated ground-state, and therefore well-approximated observable expectation values. In this sense, energy plays another role besides cost function: it is also the measure by which we judge the quality of the approximation. 

In general, this assumption does not hold and the conditions under which it fails are determined by the spectral gap~\cite{Eckart1930} and, as we will show, on the susceptibility of the observable. The fundamental consequence of this failure is that a well-converged energy does not ``certify'' or guarantee well-converged observable expectation values. The basic obstruction is familiar from perturbation theory~\cite{Kato1995}, where degeneracy generates absolute ambiguity and quasi-degeneracy generates relative ambiguity in a finite-precision computation; what matters here is the relative sizes of $\epsilon$ and the spectral gap $\Delta$. An algorithm will struggle to resolve the low-lying spectrum if $\epsilon\gtrsim\Delta$, whilst the opposite is unproblematic. We see further hints of obstruction when we consider that the variational energy is second-order in the error of the state whilst other expectation values are first-order~\cite{Helgaker2000,LuchaSchoberl2000}, leaving energy as a poor proxy for the fidelity of the state. This paper is not the first to claim as much (see, e.g.~\cite{MoraWaintal2007,Ryabinkin2019,Wang2024,LinIzmaylov2025,deLima2026} for similar recent discussions).

Thus far, ``certifiability'' hinges on the spectral gap and on whether it can be made arbitrarily small. Prior art has shown that this can be very hard to determine, and in some cases impossible~\cite{YangMills,Cubitt2022}, but in many systems of physical and chemical interest, the low-lying spectrum is quasi-degenerate, either inherently (as with the spin multiplets of open-shell molecules such as triplet CH$_2$~\cite{Niu2026}, the low-lying spin states of diradicals~\cite{Abe2013} and transition-metal clusters~\cite{Noodleman1981,Sharma2014}, the dimerised ground states of the frustrated $J_1$--$J_2$ chain~\cite{MajumdarGhosh1969,Wang2024}, the competing stripe orders of the Hubbard model~\cite{Zheng2017}) or under a deformation of the Hamiltonian (as when a bond breaks, e.g. in H$_2$~\cite{Ryabinkin2019}, first-order quantum phase transitions such as the liquid--solid transition of dipolar bosons~\cite{MoraWaintal2007}, continuous phase transitions such as that of the transverse-field Ising chain~\cite{Liu2010}). Interesting physics often lies at the boundary between gapped and gapless regimes, leaving these mechanisms of failure all the more important. 

The supposed target for accurate predictions in molecular simulation is chemical accuracy, $\epsilon_\text{chem}\approx1.6~\text{m}E_h$~\cite{Pople1999}. If these spectral gap arguments are to be believed, then a chemically accurate simulation could never certify observables for a system with $\Delta\lesssim1.6~\text{m}E_h$, which, as the examples above show, covers a great many systems. Fortunately, there is subtlety in these arguments that we shall explore. In this paper we put an observation, that large errors in observables can go unnoticed due to tightly converged energy certificates, on a formal footing. We show that, beyond the ground-to-first spectral gap, what also matters are low-lying spectral gaps and the transition matrix elements in the low-lying spectrum. Together these make up the static susceptibility of the studied observable.

\section{Certification bounds}
\label{sec:bounds}

Let $H$ have a non-degenerate ground state $\ket{\psi_0}$ with energy $E_0$, and
\be
    H\ket{\psi_n}=E_n\ket{\psi_n},\qquad E_0<E_1\leq E_2\leq\cdots,
\ee
with spectral gap $\Delta=E_1-E_0$, and let $O_{mn}=\braket{\psi_m|O|\psi_n}$ be the matrix elements of an observable $O$ in this basis such that $O_{00}$ is the ground-state expectation value. If a variational method minimises energy to within a resolution $\epsilon$, the returned state $\rho$ (in general a density matrix) satisfies $E_0\le\text{Tr}(H\rho)\le E_0+\epsilon$, so that the elements of
\be\label{eq:manifold}
    \mathcal M_\epsilon=\{\rho:E_0\le\text{Tr}(H\rho)\le E_0+\epsilon\}
\ee
are computationally indistinguishable (i.e. if the algorithm returns any $\rho\in\mathcal M_\epsilon$, the energy certifies success within an $\epsilon$ margin). We define the certification radius of an observable as the largest possible error in its expectation value compatible with that set,
\be\label{eq:radius}
    \delta_O(\mathcal{M}_\epsilon)\equiv\sup_{\rho\in\mathcal{M}_\epsilon}\lvert\text{Tr}(O\rho)-\text{Tr}(O\rho_0)\rvert,
\ee
where $\rho_0$ is the ground-state density matrix. Certification failure is the situation in which a state is ``certified'' by a well-optimised energy whilst $\delta_O$, which is compatible with that energy measurement, is disproportionately large. When that happens, the small error in energy undergoes non-linear error amplification to become a potentially large error in observable expectation value. This scale mismatch is demonstrated by way of an $\epsilon$--$\delta$ argument between the resolution $\epsilon$ and the certification radius $\delta_O$. It suffices to optimise~\eqref{eq:radius} over pure states as $\mathcal{M}_\epsilon$ is compact and convex and $\rho\mapsto\text{Tr}(O\rho)$ is affine.

We first think about this radius in terms of fidelity, $F=\lvert\braket{\psi_0|\psi}\rvert^2$. Expanding with the spectral representation of $H$,
\ba
    \epsilon\geq\braket{\psi|H-E_0|\psi}&=\sum_{n>0}(E_n-E_0)\lvert\braket{\psi_n|\psi}\rvert^2\nn\\
    &\geq\Delta\sum_{n>0}\lvert\braket{\psi_n|\psi}\rvert^2=\Delta(1-F),\nn
\ea
so that
\be\label{eq:fid}
    p\equiv1-F\le\frac{\epsilon}{\Delta},
\ee
with $p$ the population outside the ground state. This is the Eckart criterion~\cite{Eckart1930} and higher moments of $H$ can tighten it while the gap is finite~\cite{LinIzmaylov2025}. The Fuchs--van de Graaf inequality~\cite{FuchsVandeGraaf1999} then gives
\be\label{eq:trace_dist}
    T\equiv\tfrac{1}{2}\|\rho-\rho_0\|_1\le\sqrt{p}\le\sqrt{\frac{\epsilon}{\Delta}},
\ee
and H\"older's inequality converts this into a bound on any observable~\cite{Bhatia1997},
\be\label{eq:coarse}
    \delta_O\le2\|O\|_\infty\sqrt{\frac{\epsilon}{\Delta}}.
\ee
Bounds of this form go back to~\cite{Weinhold1968}, but they are weak because they see an observable only through $\|O\|_\infty$, and are therefore blind to which coherences the observable actually couples; an operator with no matrix element connecting $\ket{\psi_0}$ to the low-lying states is treated exactly like one whose weight sits entirely there. We are able to tighten the bound by accounting for this.

\begin{theorem}\label{thm:bound}
Let $O$ be bounded and $\Delta>0$. With $\Pi_\text{ex}=\mathbb I-\ket{\psi_0}\bra{\psi_0}$ the projector onto the excited states, $H_\text{ex}=\Pi_\text{ex}(H-E_0)\Pi_\text{ex}$, which is invertible on the range of $\Pi_\text{ex}$, and $O_\text{ex}=\Pi_\text{ex}(O-O_{00})\Pi_\text{ex}$,
\be\label{eq:bound}
    \delta_{O}(\mathcal M_\epsilon)\leq\epsilon\left\|H^{-\frac{1}{2}}_\text{ex}O_\text{ex}H^{-\frac{1}{2}} _\text{ex}\right\|_\infty+2\sqrt{\epsilon\, A_O(\epsilon)},
\ee
where
\be\label{eq:AO}
    A_O(\epsilon)=\sum_{n>0}\frac{\lvert O_{0n}\rvert^2}{E_n-E_0+\epsilon}.
\ee
\end{theorem}
\noindent See \app{proof1} for proof. Reaching~\eqref{eq:bound} requires a separation of the state into its ground- and excited-subspaces, $\ket{\psi}=\sqrt{1-p}\ket{\psi_0}+\sqrt p\ket u$. The first term of~\eqref{eq:bound} is carried by the population $p$ outside the ground state and is $\mathcal O(\epsilon)$, while the second is carried by the coherence between $\ket{\psi_0}$ and $\ket u$ and is $\mathcal O(\sqrt\epsilon)$. We refer to them as population and coherence terms respectively. \thm{bound} bounds $\delta_O$ from above, showing that the certification radius grows non-linearly and is affected by not just the spectral gap, but also the gaps in the low-lying spectrum and transition matrix elements. All of these factors can align to allow or disallow the possibility of disproportionate errors in observable expectation values.

We emphasise the importance of the coherence term because it dominates for small $\epsilon$ whenever $A_O\neq0$, while $A_O$ is in turn dominated by the low-lying spectrum. The resolution enters the denominators of~\eqref{eq:AO} so as to keep the coherence term physical once the gap closes. Replacing $A_O(\epsilon)$ by its $\epsilon\to0$ limit, $A_O\equiv A_O(0)$, leaves the bound correct to leading order in $\epsilon/\Delta$ but divergent as $\Delta\to0$ at fixed $\epsilon$. Subsequent loosening through $\Delta\leq E_n-E_0$ and $A_O\leq\|O\|_\infty^2/\Delta$ recovers~\eqref{eq:coarse} in the coherence term. The population term, on the other hand, has no counterpart in~\eqref{eq:coarse}, and if we similarly loosen the bound, we find a dependence on the spectral range, $\mathrm{range}(O)=\lambda_\text{max}(O)-\lambda_\text{min}(O)$:
\be\label{eq:pop_range}
    \epsilon\left\|H^{-\frac{1}{2}}_\text{ex}O_\text{ex}H^{-\frac{1}{2}}_\text{ex}\right\|_\infty\leq\frac{\epsilon}{\Delta}\,\|O_\text{ex}\|_\infty\leq\frac{\epsilon}{\Delta}\,\mathrm{range}(O).
\ee
Neither term of~\eqref{eq:bound} uses $p\leq1$, and without it both outgrow the spectrum of $O$ once $\epsilon\gg\Delta$. The population term grows linearly in $\epsilon$ without limit, and the coherence term tends to $2\sigma_O$, where $\sigma_O^2=\sum_{n>0}\lvert O_{0n}\rvert^2$ is the ground-state variance of $O$. Using $p\leq1$ directly caps the population term at $\|O_\text{ex}\|_\infty$ and the coherence term at $\sigma_O$ (see \app{proof1}), so each may be replaced by the smaller of its two bounds,
\ba\label{eq:bound_reg}
    \delta_O(\mathcal M_\epsilon)\leq{}&\min\!\left(\epsilon\left\|H^{-\frac{1}{2}}_\text{ex}O_\text{ex}H^{-\frac{1}{2}}_\text{ex}\right\|_\infty,\;\|O_\text{ex}\|_\infty\right)\nn\\
    &+\min\!\left(2\sqrt{\epsilon A_O(\epsilon)},\;\sigma_O\right),
\ea
which coincides with~\eqref{eq:bound} for $\epsilon\ll\Delta$ and is finite at every $\epsilon$ and every gap. This is the form in which the theorem should be read when the gap has closed through the resolution, as it does in every demonstration of \sec{numerics}.

To readers familiar with linear response theory~\cite{Kubo1957}, $A_O$ is not new. Under a perturbation $H\mapsto H'=H-\lambda O$,
\ba
    \braket{H'}&\mapsto E_0-\lambda O_{00}-\frac{1}{2}\chi_O\lambda^2+\mathcal O(\lambda^3),\label{eq:lr_E}\\
    \braket{O}&\mapsto O_{00}+\chi_O\lambda+\mathcal O(\lambda^2).\label{eq:lr_O}
\ea
$\chi_O$ is called the static susceptibility and $\chi_O=-\frac{d^2\braket{H'}}{d\lambda^2}=\frac{d\braket{O}}{d\lambda}$.\footnote{$\chi_O$ is computable from the ground-state alone, unlike the spectral gap which requires excited-state structure.} Explicitly, we draw the connection to $A_O$ as
\be\label{eq:chi}
    \chi_O=2\sum_{n>0}\frac{\lvert O_{0n}\rvert^2}{E_n-E_0}=2A_O,
\ee
and likewise $\chi_O(\epsilon)\equiv2A_O(\epsilon)$ is the same response with an infrared cutoff. Interestingly, the perturbations offer an alternative route to understanding the certification radius, something we hinted at in \sec{intro}. The expectation value of $H$ on the ground state of $H'$ gives $\braket{H}=E_0+\tfrac12\chi_O\lambda^2+\mathcal O(\lambda^3)$.
The energy error is second-order in $\lambda$ while the observable error is first-order, as in~\eqref{eq:lr_E} and~\eqref{eq:lr_O}. If states cannot be resolved to within $\epsilon$, $\lambda=\sqrt{2\epsilon/\chi_O}$ produces a state the energy cannot distinguish from the ground state, whose $\braket{O}$ is shifted by $\chi_O\lambda=\sqrt{2\epsilon\chi_O}$. Coming back to an earlier point,~\eqref{eq:coarse} can be strengthened simply through intuition, offering one further route to understanding the certification radius. From the perspective of $O$, the gap was never $\Delta$ but $\Delta_O\equiv\frac{2\sigma^2_O}{\chi_O}$, and if the radius saturates at $2\sigma_O$ rather than at $2\|O\|_\infty$, replacing $\|O\|_\infty\mapsto\sigma_O$ and $\Delta\mapsto\Delta_O$ sends $2\|O\|_\infty\sqrt{\epsilon/\Delta}\mapsto\sqrt{2\epsilon\chi_O}$. Once again we recover \thm{bound} to leading order, and with static susceptibility in mind the theorem may be reframed.
\begin{corollary}\label{cor:chi}
With $\chi_O(\epsilon)$ the static susceptibility~\eqref{eq:chi} regulated at the resolution,
\ba\label{eq:bound_chi}
    \delta_{O}(\mathcal M_\epsilon)&\leq\epsilon\left\|H^{-\frac{1}{2}}_\text{ex}O_\text{ex}H^{-\frac{1}{2}}_\text{ex}\right\|_\infty+\sqrt{2\epsilon\chi_O(\epsilon)}\nn\\
    &\leq\epsilon\left\|H^{-\frac{1}{2}}_\text{ex}O_\text{ex}H^{-\frac{1}{2}}_\text{ex}\right\|_\infty+\sqrt{2\epsilon\chi_O}.
\ea
\end{corollary}
\noindent \thm{bound} and this corollary bound $\delta_O$ from above. We now show that the radius is also bounded below by the same susceptibility.

\begin{theorem}\label{thm:converse}
Let $\chi_O>0$ and
\be\nn
    \bar\Delta_O=\frac{\sum_{n>0}\lvert O_{0n}\rvert^2(E_n-E_0)^{-1}}{\sum_{n>0}\lvert O_{0n}\rvert^2(E_n-E_0)^{-2}},
\ee
so that $\Delta\;\leq\;\bar\Delta_O\;\leq\;\Delta_O$. For $\epsilon\le\bar\Delta_O/2$,
\be\label{eq:converse}
    \delta_O(\mathcal M_\epsilon)\geq\sqrt{2\epsilon\chi_O}\,\sqrt{1-\epsilon/\bar\Delta_O}.
\ee
\end{theorem}
\noindent See \app{proof2} for proof. As for \thm{bound}, this follows from a decomposition $\ket{\psi}=\sqrt{1-p}\ket{\psi_0}\pm\sqrt p\ket u$ with $p=\epsilon/\bar\Delta_O$. Together, the two theorems sandwich the radius.

\begin{corollary}\label{cor:sandwich}
For $\epsilon\le\bar\Delta_O/2$,
\ba\label{eq:sandwich}
    \sqrt{2\epsilon\chi_O}\,\sqrt{1-\epsilon/\bar\Delta_O}&\leq\delta_O(\mathcal M_\epsilon)\nn\\
    &\leq\epsilon\left\|H^{-\frac{1}{2}}_\text{ex}O_\text{ex}H^{-\frac{1}{2}}_\text{ex}\right\|_\infty+\sqrt{2\epsilon\chi_O(\epsilon)},
\ea
so that $\delta_O(\mathcal M_\epsilon)=\sqrt{2\epsilon\chi_O}\,[1+\mathcal O(\sqrt\epsilon)]$ whenever $\chi_O>0$.
\end{corollary}
\noindent To leading order in $\epsilon$, then, an observable is uncertifiable precisely to the extent that it is susceptible. In other words, energy alone can never guarantee a susceptible observable expectation value to better than $\sqrt{2\epsilon\chi_O}$, however the state was prepared. It does not, however, prove that the error must be that large, only that it may be, and that an energy benchmark cannot tell.

Let us now turn to symmetry considerations. Symmetry is able to restore certifiability through three key mechanisms. First and simplest is that, since $O_{0n}(E_n-E_0)=\bra{\psi_0}[H,O]\ket{\psi_n}$, if $O$ is a conserved quantity, then the coherence term vanishes and the dominant scaling improves from $\sqrt{\epsilon/\Delta}$ to $\epsilon/\Delta$ (the population term).
Second, let $[H,S]=0$ for a symmetry operator $S$, which partitions the Hilbert space via symmetry-sector projectors $\Pi_{s_n}$. A symmetry-adapted ansatz confined to the ground-state sector sees only the restricted set
\be\label{eq:manifold_sym}
    \mathcal M_{\epsilon,s_0}=\left\{\rho:\rho=\Pi_{s_0}\rho\Pi_{s_0},\;E_0\leq\text{Tr}(H\rho)\leq E_0+\epsilon\right\},
\ee
and \thm{bound} then holds with the in-sector gap
\be
    \Delta_{s_0}=\min_{\substack{n>0\\s_n=s_0}}(E_n-E_0)\;\geq\;\Delta.
\ee
Approximate symmetry breaking is characterised by a small leakage $\ell(\psi)=\|(\mathbb I-\Pi_{s_0})\ket{\psi}\|^2$, which reintroduces coherence such that
\ba\label{eq:leakage}
    \delta_{O}\leq{}&\delta_{O}(\mathcal M_{\epsilon/(1-\ell),s_0})+\ell\,\mathrm{range}(O)\nn\\
    &+2\sqrt{\ell(1-\ell)}\,\bigl\|(\mathbb I-\Pi_{s_0})\,O\,\Pi_{s_0}\bigr\|_\infty.
\ea
Third, combining the two, a selection rule, $\Pi_{s_0}O\ket{\psi_0}=0$, enforces $O_{0n}=0$ for all $n>0$ with $s_n=s_0$ such that the coherence channel vanishes.

Several gaps have appeared throughout this section: $\Delta$, $\Delta_O$, $\bar\Delta_O$, and $\Delta_{s_0}$. The studies that follow show which of them governs certifiability and in what contexts.

\section{Numerical studies}
\label{sec:numerics}

To test the claims of the previous section, we now study three systems: the transverse-field Ising chain, a pair of weakly coupled hydrogen chains, and lithium fluoride, under three different implementations of the variational principle to show how universal certification failure is. In each, the resolution is fixed to achieve tight convergence in energy and then a deformation of the Hamiltonian closes the spectral gap. We shall see that the observable error becomes large once the two cross.\footnote{The error need not reach the radius: the states that saturate it are reachable by the optimiser, but nothing guarantees that it finds them.} Each study pairs an example that fails with one, run under the same protocol, that does not, and each demonstrates a different mechanism of failure. Between them they exercise both terms of \thm{bound} and each of the three ways in which symmetry restores certifiability (see \tab{studies} for an overview). Computational details of how the three protocols are implemented can be found in \app{details}.

\begin{table*}[t]
\centering
\setlength{\tabcolsep}{9pt}
\renewcommand{\arraystretch}{1.25}
\begin{tabular}{@{}llllll@{}}
\toprule
Method & System & $O$ & Closing gap & Failing term & Protection \\
\midrule
VQE & Ising chain & $M$ & parity doublet, $h\downarrow$ & coherence, $2|M_{01}|\sqrt{\epsilon/\Delta}$ & parity sector, $M$ odd\\
DMRG & H$_5$ chains & $\braket{S^2}$ & singlet--triplet, $d\uparrow$ & population, $\epsilon/\Delta$ & spin sector, gap $\Delta_{s_0}$\\
KQD & LiF & $\mu$ & ionic--covalent, $r\uparrow$ & coherence, $2|\mu_{01}|\sqrt{\epsilon/\Delta}$ & detuned, gap held open\\
\bottomrule
\end{tabular}
\caption{Overview of the three studies. In each, a deformation closes the gap $\Delta$ between the two lowest states. The failing term is the leading term of \thm{bound} that carries the observable error of the failing example.}
\label{tab:studies}
\end{table*}

With numerics, rather than rely on the analytical bounds of \sec{bounds}, we can map out exactly the certification radius. To find the radius is to find $\max_\rho\text{Tr}(O\rho)$ subject to the constraint $\text{Tr}(H\rho)\leq E_0+\epsilon$. Using a Lagrange multiplier $1/\lambda$, the optimisation becomes
\be\label{eq:opt}
    \max_\rho\left[\text{Tr}(O\rho)-\frac{1}{\lambda}\bigl(\text{Tr}(H\rho)-E_0-\epsilon\bigr)\right].
\ee
If $\epsilon$ is the difference between the variational energy $\braket{H}$ and the exact energy $E_0$ for every computation then we simply re-express~\eqref{eq:opt} as $\frac{\braket{H}-E'(\lambda)}{\lambda}$ where $E'(\lambda)$ is the exact ground-state energy of $H'$. Every $\lambda>0$ gives a ceiling on $\braket{O}$ and we choose the $\lambda$ that minimises it. The same with $\lambda<0$ gives the floor~\cite{CohenFeldmann1971}. The interval between floor and ceiling is the certification region shaded in the figures, and its largest distance from $O_{00}$ is the radius.

\emph{VQE.} We perform classical simulation of the variational quantum eigensolver (VQE)~\cite{Peruzzo2014} on an open transverse-field Ising chain of $L=12$ sites~\cite{Pfeuty1970},
\be\label{eq:tfim}
    H=-J\sum_iZ_iZ_{i+1}-h\sum_iX_i,
\ee
where $X_i$ and $Z_i$ are the Pauli operators on site $i$ and energies are in units of the Ising coupling $J$. We lower the transverse field $h$ from the paramagnet into the ordered phase. $H$ commutes with the parity $P=\prod_iX_i$ at every $h$. In the ordered phase the ground- and first-excited states form a doublet of opposite parity whose splitting $\Delta$ closes exponentially as $h$ falls, while the gap $\Delta_s$ inside the ground state's sector stays of order one. The observable we consider is the magnetisation $M=L^{-1}\sum_i\braket{Z_i}$. It is odd under $P$, so it vanishes in the ground state, but any superposition of the two members of the doublet is magnetised, with a sign an energy measurement cannot choose.

\begin{figure}[t]
    \centering
    \includegraphics[width=\columnwidth]{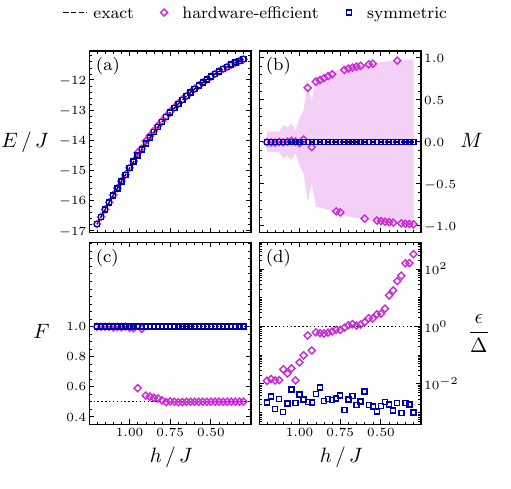}
    \caption{VQE on the transverse-field Ising chain~\eqref{eq:tfim}, $L=12$, as the transverse field $h$ is lowered; energies and $h$ in units of the Ising coupling $J$. (a) Energy. (b) Magnetisation, zero in the ground state. The shaded region sweeps out the exact certification radius. For the parity-preserving ansatz the region vanishes by symmetry protection. (c) Fidelity with the ground state; one half and one dotted. (d) Energy error over the gap each ansatz sees: $\Delta$ for the hardware-efficient ansatz and $\Delta_s$ for the parity-preserving one; unity dotted.}
    \label{fig:tfim}
\end{figure}

We consider two parameterised circuits of the same depth as variational ans\"atze. The first is a hardware-efficient ansatz~\cite{Kandala2017}, layers of single-qubit rotations followed by a ladder of entangling gates, which does not conserve $P$ and so sees the whole doublet. The second is a parity-preserving ansatz of the Hamiltonian-variational type~\cite{Wecker2015,Wiersema2020}, with one angle per bond and per site in every layer, which commutes with $P$ and never leaves the ground state's sector. Both are optimised identically from independent random starts, and while the gap is open both energy errors stay below $0.015$ (\fig{tfim}[a]). Their magnetisations do not agree. As the splitting closes, the hardware-efficient runs settle on one member of a broken-symmetry pair, with the sign set by the random start. At $h=0.9$, where the splitting is $0.13$, the run shown has $|M|=0.72$ and a fidelity near one half, at an energy error of $0.08$, about half the splitting. By the end of the sweep $|M|=0.98$, at an energy error three hundred times the splitting. The parity-preserving ansatz returns $M=0$ at every $h$. Its resolution is the same, but the gap it sees is $\Delta_s\geq0.76$, more than two orders of magnitude above its energy error, and for an odd observable the protection is complete, since $M$ has no matrix element inside a sector and the region collapses to zero.

\emph{DMRG.}\label{sec:hchain} We run the density-matrix renormalisation group (DMRG)~\cite{White1992,Schollwoeck2011} on two H$_5$ chains in 6-31G, with $1.8\,a_0$ between neighbouring atoms, placed end to end and pulled apart to a separation $d$. Each chain carries one unpaired electron, so the pair is a model diradical. Its ground state is a singlet and its gap is the singlet--triplet splitting $J$ (not to be confused with the Ising coupling), which is set by tunnelling between the chains and closes exponentially as $d$ grows. The error of DMRG at a fixed bond dimension $D$ is set instead by the correlation within each chain, and does not change with $d$. The observable we consider is the total spin $\braket{S^2}$, which is zero in the ground state. It commutes with $H$, such that the coherence term of \thm{bound} vanishes and the region grows linearly in $\epsilon/J$.

\begin{figure}[t]
    \centering
    \includegraphics[width=\columnwidth]{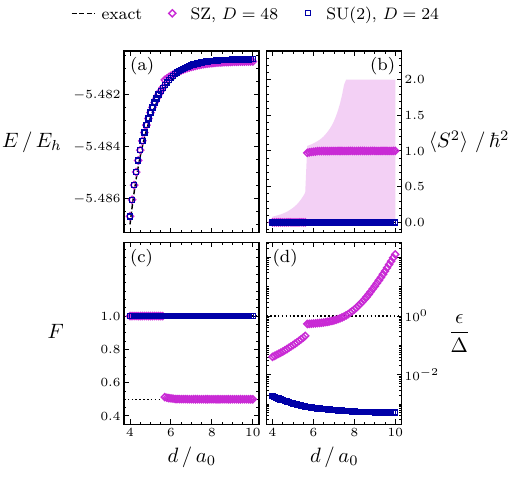}
    \caption{DMRG on two H$_5$ chains in 6-31G as their separation $d$ grows; SZ DMRG at $D=48$ and SU(2) DMRG at $D=24$. (a) Energy. (b) Total spin $\braket{S^2}$, zero in the ground state. The shaded region sweeps out the exact certification radius. For SU(2) DMRG the region vanishes by symmetry protection. (c) Fidelity with the ground state; one half and one dotted. (d) Energy error over the gap each formulation sees: $J$ for SZ DMRG and $\Delta_{s_0}$ for SU(2) DMRG; unity dotted.}
    \label{fig:dmrg}
\end{figure}

We consider two formulations of the method at matched resolution. The first conserves only $S_z$ (SZ DMRG, $D=48$) and so sees the triplet. The second is spin adapted (SU(2) DMRG~\cite{Sharma2012}, $D=24$) and never leaves the singlet sector. Both are run identically from independent random starts, and while the gap is open both energy errors stay below $0.31\,\text{m}E_h$ (\fig{dmrg}[a]). Their spins do not agree. At $d=5.7\,a_0$, where $J=139\,\text{cm}^{-1}$, a typical diradical coupling, every SZ start switches to the broken-symmetry state, with one spin localised on each chain, $\braket{S^2}=1$ and a fidelity of one half, at an energy error of $0.34\,\text{m}E_h$, about half the gap and five times below chemical accuracy. By the end of the sweep $\braket{S^2}=1$, at an energy error a hundred times the gap. The singlet of two distant spins carries an extra unit of entanglement between the chains, which a finite bond dimension must pay for, while breaking the symmetry costs only $J/2$~\cite{Noodleman1981}. Up to $6.3\,a_0$ a lower SZ singlet still exists at $D=48$, but no random start finds it; beyond that the broken-symmetry state is also the lower in energy, so even a comparison of the two calculations would choose it. Raising $D$ moves the switch out but does not remove it: to $J=7\,\text{cm}^{-1}$ at $D=96$, and to $2\,\text{cm}^{-1}$ at $D=192$, where the energy is converged to $5\,\mu E_h$. SU(2) DMRG returns $\braket{S^2}=0$ at every $d$. Its resolution is the same, but the gap it sees is the singlet--singlet gap $\Delta_{s_0}\geq170\,\text{m}E_h$, an excitation within the chains, more than two orders of magnitude above its energy error, and inside its sector the region collapses to zero.

\emph{KQD.}\label{sec:lif} We perform classical simulation of Krylov quantum diagonalisation (KQD)~\cite{Cortes2022} on lithium fluoride in SA-CASSCF(6,6)/aug-cc-pVDZ. Unlike VQE and DMRG, where the variational principle is explicit, KQD invokes it only at the point of classical diagonalisation via the Rayleigh--Ritz method. Errors from the quantum computer pass through to the output via noisy matrix elements. At the ionic--covalent avoided crossing of LiF, the two lowest states differ in dipole by tens of debye~\cite{Werner1981,Ushakov2025}. A uniform electric field along the bond, $H(r,F)=H_0(r)-F\mu$, moves the crossing to any separation $r$. On the resonance line $F=F_c(r)$ the two states are held at the crossing, and the gap closes as $r$ grows; on the detuned line the field is shifted just enough to hold the gap open. The observable we consider is the dipole $\mu$, whose static susceptibility is the polarisability $\alpha$.

\begin{figure}[t]
    \centering
    \includegraphics[width=\columnwidth]{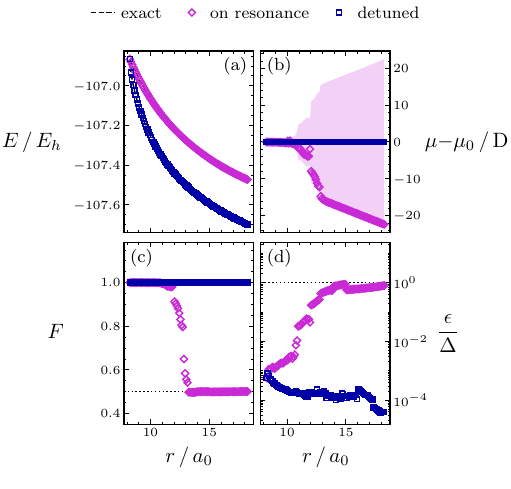}
    \caption{KQD on LiF in SA-CASSCF(6,6)/aug-cc-pVDZ as the separation $r$ grows, on resonance and detuned. (a) True energy $\braket{\psi|H|\psi}$. (b) Dipole relative to its exact value $\mu_0$. The shaded region sweeps out the exact certification radius. (c) Fidelity with the ground state; one half and one dotted. (d) True energy error over the gap; unity dotted.}
    \label{fig:sqdrift}
\end{figure}

\begin{figure}[t]
    \centering
    \includegraphics[width=\columnwidth]{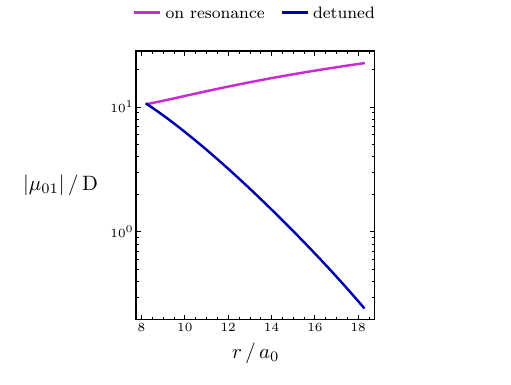}
    \caption{The transition dipole $|\mu_{01}|$ between the two lowest states of LiF as the separation $r$ grows, on resonance and detuned.}
    \label{fig:mu01}
\end{figure}

We run the same protocol on both lines, which meet at $8.25\,a_0$ and differ only in the field, and judge every run by the true energy of the state it returns (\fig{sqdrift}[a]), since the reported eigenvalue can lie below $E_0$. While the gap is open both energy errors stay below $0.06\,\text{m}E_h$. On resonance the dipole fails before the fidelity does, through the coherence term: at $r=11\,a_0$ a run with a fidelity of $0.995$ already has a dipole $1.7$ debye wrong. As the gap closes further, the direction of the overlap matrix that separates the pair falls below the noise threshold and is discarded~\cite{Epperly2022,Kirby2024}, and the returned state becomes an equal superposition of the two, which costs half the gap and moves the dipole by the transition dipole $|\mu_{01}|$ (\fig{mu01}). At $r=13.25\,a_0$, where the gap is $2.7\,\text{m}E_h$, the fidelity reaches one half and the dipole is $16$ debye wrong, at an energy error of $1.4\,\text{m}E_h$, about half the gap and just below chemical accuracy. By the end of the sweep the dipole is $22$ debye wrong, at an energy error comparable to the gap. With exact matrices the same Krylov space resolves the pair at every $r$. Detuned, the same protocol returns the right dipole to within $0.04$ debye. Its resolution is the same, but the gap is held at $22\,\text{m}E_h$, three orders of magnitude above its energy error, and the two states barely mix, so $|\mu_{01}|$ collapses some fortyfold and $\alpha$, which rises four hundredfold along the resonance line, stays small.

\section{Discussion}
\label{sec:disc}

This entire exercise has been about determining when we can trust variational computation and when we cannot, which, as we have shown, is a property of the method, of the system, and most importantly, of the observable. If only the ground-state energy is needed then none of the above applies, and the variational principle certifies it directly. Increasing the accuracy (reducing $\epsilon$) of a method is not always possible but we have shown that even when it is, it is not necessarily sufficient, and that there exist fundamental obstructions to guarantees on observable expectation values.

One consequence is that accuracy targets belong to observables rather than to energies. In other words, to know $\braket{O}$ to within $\delta_\ast$ one needs, at leading order,
\be\label{eq:target}
    \epsilon\lesssim\frac{\delta_\ast^2}{2\chi_O}.
\ee
Ensuring that $\epsilon$ reaches chemical accuracy is not in itself informative about $\braket{O}$, not because the energy is unimportant, but because it is a single energy scale, and certification requires comparing several: the resolution, the gap, and through $\chi_O$, the transition elements that couple the ground-state to the low-lying spectrum. 

Setting the transition elements aside, we can say something interesting about certifiability in the asymptotic limit through the ratio $r=\lim_{L\rightarrow\infty}\epsilon/\Delta$. At a critical point, for example, the spectral gap closes as a power of the system size, $\Delta\sim L^{-z}$, where $z>0$ is some critical exponent~\cite{Sachdev2011}. We can also attribute to every algorithm some effective resolution, $\epsilon_\text{eff}$, which weds its various sources of error and noise and scales with system size. If $r$ diverges, then the algorithm-system pair undergoes certification failure in the asymptotic limit.\footnote{This excludes the unlikely possibility that the transition elements simultaneously vanish at a rate which allows the radius to remain finite.} In fact, it would take $\epsilon_\text{eff}\lesssim L^{-z}$ in order for certification to remain a possibility; in other words the error would need to shrink as the system grows and at a sufficient rate. This is plainly impossible for VQE where the dominant sources of error, statistical shot variance and gate noise, both worsen as the norm of the Hamiltonian and the depth of the circuit grow with system size~\cite{Loaiza2023,Xia2021}, and in the absence of convergence guarantees~\cite{McClean2018Barren}. DMRG and KQD are more interesting cases. Keeping $\epsilon\lesssim\Delta$ in KQD requires a Krylov dimension and a shot count that grow as powers of $L$, which is in principle possible, but only from an initial state whose overlap does not itself decay exponentially with system size~\cite{Epperly2022,Kirby2024}. For DMRG, the weight discarded at a fixed bond dimension grows with the entanglement of the state, which at criticality grows only with the logarithm of the length~\cite{CalabreseCardy2004}, so a bond dimension growing polynomially in $L$ can keep pace with a gap closing as $L^{-\gamma}$~\cite{VerstraeteCirac2006,Tagliacozzo2008}. It cannot, however, keep pace with a gap that closes exponentially, as with the weakly coupled chains of \sec{hchain}.

The problem we present is confined neither to ground-states, nor to local observables. The variational principle admits generalisations to excited states~\cite{Hylleraas1930,MacDonald1933} and we present arguments about certification of excited states in \app{excited}. We are also able to consider non-observable quantities such as the half-chain entanglement entropy for which the Fannes--Audenaert inequality gives a certification radius $\delta_S\le T\log(d_A-1)+h(T)$, with $T\le\sqrt{\epsilon/\Delta}$, $h$ the binary entropy and $d_A$ the dimension of the half-chain~\cite{Fannes1973,Audenaert2007}.

Much of the relevant information a quantum chemist might require lies on the potential energy surface itself, where large errors in observable expectation values are of no consequence so long as the energy error remains small. Elsewhere, these results call for caution, and for a few solutions. The first is to benchmark beyond the energy; $\chi_O$ is computable from the ground-state alone, as a first difference of $\braket{O}$ under $H-\lambda O$~\cite{CohenRoothaan1965}, and with $\epsilon$ it gives the leading-order radius directly. The second is to use a symmetry-adapted ansatz wherever a symmetry is exact, since a conserved observable then loses its coherence term and an odd one its entire radius. The third is to track $\epsilon/\Delta$ across any deformation known to close a gap, rather than $\epsilon$ alone.

\acknowledgments{The author thanks David P. Tew and Lewis W. Anderson for helpful feedback and Man Hei Cheng for early encouragement. The author acknowledges the assistance of various ChatGPT and Claude models in the preparation of this manuscript and assumes full responsibility for its scientific content and integrity. This work was supported by the EPSRC and IBM Research Europe through an Industrial Doctoral Landscape Award.}

\bibliography{bib}

\newpage
\widetext

\makeatletter
\renewcommand\section{\@startsection{section}{1}{\z@}%
  {2.6ex \@plus 1ex \@minus .2ex}{1.2ex \@plus .2ex}%
  {\normalfont\normalsize\bfseries\centering}}
\makeatother

\appendix
\input{appendices.tex}

\end{document}

%% file: appendices.tex
\begin{center}
    \textbf{\large Appendices}
\end{center}

\section{Proof of \thm{bound}}
\label{app:proof1}

As $\mathcal M_\epsilon$ is convex and $\rho\mapsto\text{Tr}(O\rho)$ is affine, it is enough to consider a pure state $\ket\psi$ with $E_0\leq\braket{\psi|H|\psi}\leq E_0+\epsilon$. Let $p=1-\lvert\braket{\psi_0|\psi}\rvert^2$ be its weight outside the ground state and, choosing the global phase so that $\braket{\psi_0|\psi}\geq0$, write
\be\label{eq:decomp}
    \ket\psi=\sqrt{1-p}\,\ket{\psi_0}+\sqrt p\,\ket u,\qquad \Pi_\text{ex}\ket u=\ket u,\quad\braket{u|u}=1.
\ee
Because $(H-E_0)\ket{\psi_0}=0$, only $\ket u$ costs energy,
\be\label{eq:proof_energy}
    \braket{\psi|H-E_0|\psi}=p\,\braket{u|H_\text{ex}|u}\leq\epsilon,
\ee
and, since $\ket u\perp\ket{\psi_0}$, the shift in $\braket{O}$ splits into two terms,
\be\label{eq:proof_split}
    \braket{\psi|O|\psi}-O_{00}=\underbrace{p\,\braket{u|O_\text{ex}|u}}_\text{population}+\underbrace{2\sqrt{p(1-p)}\;\text{Re}\braket{\psi_0|O|u}}_\text{coherence},
\ee
each of which we bound against~\eqref{eq:proof_energy}.

\emph{Population.} Since $\Delta>0$, $H_\text{ex}$ is positive and invertible on the excited subspace. With $\ket v=H_\text{ex}^{1/2}\ket u$, so that $\braket{v|v}=\braket{u|H_\text{ex}|u}$,
\be\nn
    \bigl\lvert p\braket{u|O_\text{ex}|u}\bigr\rvert=p\,\bigl\lvert\braket{v|H^{-\frac12}_\text{ex}O_\text{ex}H^{-\frac12}_\text{ex}|v}\bigr\rvert\leq\epsilon\left\|H^{-\frac12}_\text{ex}O_\text{ex}H^{-\frac12}_\text{ex}\right\|_\infty,
\ee
by~\eqref{eq:proof_energy}.

\emph{Coherence.} Write $H_\text{ex}+\epsilon$ for $H_\text{ex}+\epsilon\Pi_\text{ex}$, insert $(H_\text{ex}+\epsilon)^{-1/2}(H_\text{ex}+\epsilon)^{1/2}$ between $O$ and $\ket u$, and apply the Cauchy--Schwarz inequality~\cite{HoffmannOstenhof1975}:
\be\nn
    \lvert\braket{\psi_0|O|u}\rvert\leq\bigl\|(H_\text{ex}+\epsilon)^{-\frac12}\Pi_\text{ex}O\ket{\psi_0}\bigr\|\;\bigl\|(H_\text{ex}+\epsilon)^{\frac12}\ket u\bigr\|=\sqrt{A_O(\epsilon)}\,\sqrt{\braket{u|H_\text{ex}|u}+\epsilon},
\ee
since the first norm squared is~\eqref{eq:AO}. The resolution in its denominators comes from this step: inserting $H_\text{ex}^{-1/2}H_\text{ex}^{1/2}$ instead would give $A_O(0)$, which diverges as $\Delta\to0$. By~\eqref{eq:proof_energy}, $p(1-p)\bigl(\braket{u|H_\text{ex}|u}+\epsilon\bigr)\leq(1-p)\epsilon+p\epsilon=\epsilon$, so
\be\nn
    2\sqrt{p(1-p)}\,\lvert\braket{\psi_0|O|u}\rvert\leq2\sqrt{\epsilon\,A_O(\epsilon)}.
\ee
Adding the two bounds gives~\eqref{eq:bound}. Neither bound uses $p\leq1$, and without it both outgrow the spectrum of $O$ once $\epsilon\gg\Delta$: the population term grows linearly in $\epsilon$ and the coherence term tends to $2\sigma_O$. Using $p\leq1$ directly, the population term is at most $\|O_\text{ex}\|_\infty$ and, since $2\sqrt{p(1-p)}\leq1$ and $\lvert\braket{\psi_0|O|u}\rvert\leq\|\Pi_\text{ex}O\ket{\psi_0}\|=\sigma_O$, the coherence term is at most $\sigma_O$. Taking the smaller bound on each term gives~\eqref{eq:bound_reg}.\hfill$\square$

\section{Proof of \thm{converse}}
\label{app:proof2}

The Cauchy--Schwarz step of \app{proof1} is saturated when $\ket u$ points along the first-order response of the ground state to a field coupled to $O$, and displacing the ground state along that direction by as much as the energy allows gives the lower bound. For $n>0$ write $w_n=E_n-E_0$ and $c_n=\lvert O_{0n}\rvert^2$, and let
\be\label{eq:AB}
    A_O=\sum_{n>0}\frac{c_n}{w_n}=\frac{\chi_O}{2},\qquad B_O=\sum_{n>0}\frac{c_n}{w_n^2},\qquad \bar\Delta_O=\frac{A_O}{B_O}.
\ee
Both sums are positive because $\chi_O>0$. As a mean of the $w_n$ weighted by $c_n/w_n^2$, $\bar\Delta_O\geq\Delta$, and as $A_O^2\leq\sigma_O^2B_O$ by the Cauchy--Schwarz inequality, $\bar\Delta_O\leq\sigma_O^2/A_O=\Delta_O$. Under $H\mapsto H-\lambda O$ the ground state moves at first order along $H_\text{ex}^{-1}\Pi_\text{ex}O\ket{\psi_0}=\sum_{n>0}(O_{0n}/w_n)\ket{\psi_n}$, whose norm squared is $B_O$. Normalising it,
\be\nn
    \ket u=\frac{H_\text{ex}^{-1}\Pi_\text{ex}O\ket{\psi_0}}{\sqrt{B_O}},\qquad \braket{\psi_0|O|u}=\frac{A_O}{\sqrt{B_O}},\qquad \braket{u|H_\text{ex}|u}=\bar\Delta_O.
\ee
The states $\ket{\psi_\pm}=\sqrt{1-p}\,\ket{\psi_0}\pm\sqrt p\,\ket u$ have energy error $p\,\bar\Delta_O$, so at $p=\epsilon/\bar\Delta_O\leq\tfrac12$ both lie in $\mathcal M_\epsilon$, and by~\eqref{eq:proof_split}
\be\nn
    \braket{\psi_\pm|O|\psi_\pm}-O_{00}=x\pm y,\qquad x=p\,\braket{u|O_\text{ex}|u},\qquad y=2\sqrt{p(1-p)}\,\frac{A_O}{\sqrt{B_O}}\geq0.
\ee
Whatever the sign of $x$, one of the two has $\lvert x\pm y\rvert\geq y$, so $\delta_O(\mathcal M_\epsilon)\geq y$, and substituting $p$,
\be\nn
    y=2\sqrt{\epsilon A_O}\,\sqrt{1-p}=\sqrt{2\epsilon\chi_O}\,\sqrt{1-\epsilon/\bar\Delta_O},
\ee
which is~\eqref{eq:converse}. Since $\mathcal M_\epsilon$ grows with $\epsilon$, for $\epsilon>\bar\Delta_O/2$ the radius is at least its value at $\bar\Delta_O/2$.\hfill$\square$

\section{Excited states}
\label{app:excited}

For an interior eigenvalue $E_k$ the energy certifies nothing, because weight below and above $E_k$ cancels in $\braket{H}$; the certificate is instead the variance~\cite{Temple1928}. Let $E_k$ be non-degenerate, with $Q_k=\mathbb I-\ket{\psi_k}\bra{\psi_k}$, $K_k=Q_k(H-E_k)Q_k$ and $O_{k,\text{ex}}=Q_k(O-O_{kk})Q_k$. The argument of \app{proof1}, with $(H-E_k)^2$ in place of $H-E_0$, gives for every $\rho$ with $\text{Tr}[(H-E_k)^2\rho]\leq\nu$
\ba\label{eq:excited_bound}
    \left\lvert O_{00}-O_{kk}\right\rvert\leq{}&\nu\left\||K_k|^{-1}O_{k,\text{ex}}|K_k|^{-1}\right\|_\infty\nn\\
    &+2\sqrt\nu\left(\sum_{m\neq k}\frac{|O_{mk}|^2}{(E_m-E_k)^2+\nu}\right)^{1/2}.
\ea
The population term alone is the overlap bound in~\cite{Blau1973}. The denominators are squared where~\eqref{eq:bound} has them linear: certification is one-sided at the spectral edge and two-sided in the interior, and two-sidedness costs one more power of the gap.

\section{Computational details}
\label{app:details}

Each algorithm is run as an ensemble of independent random starts, never warm-started, with every member kept. At each point of a sweep the runs within a factor of two of the best energy error are tied, and the figures show the tied run whose observable is furthest from its exact value. The exact certification region at the energy of that run comes from the ground-state energies of $H\mp\lambda O$ on a logarithmic grid of $\lambda$ or, for a conserved $O$, from the lowest energy in each eigenspace of $O$. Integrals and reference calculations use PySCF~\cite{Sun2020}, and DMRG uses \textsc{block2}~\cite{Zhai2023}.

\emph{VQE (\fig{tfim}).} The chain~\eqref{eq:tfim} has $L=12$ open sites, at $37$ values of $h$ from $1.2$ to $0.3$, and the reference is exact diagonalisation within each parity sector, with $131$ values of $\lambda$ from $10^{-9}$ to $10^4$. Both circuits have $12$ layers. The hardware-efficient ansatz applies $R_X$ and $R_Y$ to every qubit followed by a ladder of CZ gates, with $288$ angles drawn uniformly from $[0,2\pi)$. The parity-preserving ansatz prepares $\ket{+}^{\otimes L}$ and then applies $\exp(-it_bZ_iZ_{i+1})$ on every bond and $\exp(-is_iX_i)$ on every site, with $276$ angles drawn uniformly from $[-0.01,0.01]$. States and gradients come from a statevector simulation with adjoint differentiation. Each start is optimised with Adam at a step size of $0.03$ until the energy changes by less than $10^{-9}$ or for at most $800$ steps, with $24$ starts per $h$ and per ansatz.

\emph{DMRG (\fig{dmrg}).} The two H$_5$ chains are collinear, at $61$ separations $d$ from $4$ to $10\,a_0$, in L\"owdin-orthogonalised 6-31G orbitals ordered along the axis ($20$ orbitals, $10$ electrons). Two-site DMRG is run with $S_z=0$ (SZ) or $S=0$ (SU(2)) symmetry from random matrix-product states, with $6$ starts per $d$. Each run makes $30$ sweeps: four at each bond dimension in $\{20,50,100,200\}$ below $D$ with noise $10^{-4}$, then the rest at $D$ with noise $10^{-5}$, $10^{-6}$ and zero in roughly equal thirds, to an energy tolerance of $10^{-12}\,E_h$. The reference is SU(2) DMRG at $D=200$ for the two lowest singlets and the lowest triplet, and at $D=96$ for $S=2,\dots,5$; the region is the upper convex hull of $(E_S-E_0,\,S(S+1))$, and fidelities are overlaps with the singlet reference. The runs at $D=96$ and $192$ use SZ DMRG with $3$ starts every $0.2\,a_0$, and place the switch at the first separation at which $\braket{S^2}>0.5$.

\emph{KQD (\fig{sqdrift}).} LiF is described by SA-CASSCF(6,6)/aug-cc-pVDZ over the two lowest $^1\Sigma^+$ states, at $101$ separations $r$ from $8.25$ to $18.25\,a_0$. At each $r$ the orbitals are optimised once, at the field predicted by a two-state diabatic model, and held fixed while the field is varied. $H$ and $\mu$ are built as dense $400\times400$ matrices over the active space, with a penalty $1\,E_h\times S^2$ that keeps the two lowest states singlet, and are diagonalised exactly for the reference, with $181$ values of $\lambda$ from $10^{-7}$ to $10^2$. $F_c(r)$ minimises the gap between the two singlets, and the detuned field holds it at $21.78\,\text{m}E_h$ with the ionic state below. KQD starts from the closed-shell reference determinant, with Krylov states at $t_k=5k$ a.u., Krylov dimensions $D_\text{K}\in\{4,6,8,12,16,24\}$ and exact time evolution. The projected Hamiltonian and overlap matrices receive Hermitian Gaussian noise of standard deviation $\|\tilde H\|/\sqrt N$ and $1/\sqrt N$ for $N\in\{10^5,10^7,10^9\}$ shots, and overlap eigenvalues below $3/\sqrt N$ of the largest are discarded before the generalised eigenproblem is solved. Each line has $64$ starts per $(D_\text{K},N,r)$, and runs are tied and drawn by their true energy.